\documentclass[twocolumn,tighten,times,twocolappendix]{aastex701}
\usepackage[T1]{fontenc}
\usepackage[utf8]{inputenc}
\DeclareUnicodeCharacter{2009}{\nobreakspace}
\usepackage[dvipsnames]{xcolor}
\usepackage{graphicx}
\usepackage{xspace}
\usepackage{amssymb, amsmath, mathtools}
\usepackage{bm}
\usepackage{microtype}
\usepackage[english]{babel}
\usepackage{hyperref}
\usepackage{times}
\usepackage[varg]{txfonts}

\newcommand{\IACS}{\affiliation{Indian Association for the Cultivation of Science, Kolkata - 700032, India }}
\newcommand{\ICTS}{\affiliation{International Centre for Theoretical Sciences, Tata Institute of Fundamental Research, Bengaluru - 560089, India}}
\newcommand{\IUCAA}{\affiliation{The Inter-University Centre for Astronomy and Astrophysics (IUCAA), Post Bag 4, Ganeshkhind, Pune 411007, India}}

\shorttitle{Microlensing Biases in Ringdown Tests of GR}
\shortauthors{Dasan, Mishra and Ganguly}

\graphicspath{{./}{figs/}}

\begin{document}

\title{Biases in Ringdown Tests of General Relativity from Gravitational-Wave Microlensing}

\author[0009-0005-1818-7285]{Sreelakshmi Mundakkal Dasan}
\email{spssmd3290@iacs.res.in} 
\IACS
\IUCAA

\author[orcid=0000-0002-2580-2339]{Anuj Mishra}
\email{anuj.mishra@icts.res.in}
\ICTS

\author[orcid=0000-0001-7394-0755]{Apratim Ganguly}
\email{apratim@iucaa.in}
\IUCAA

\date{\today}
 
\begin{abstract}
Gravitational-wave (GW) ringdown signals provide a powerful probe of the strong-field regime of general relativity (GR) through black hole spectroscopy and tests of the Kerr hypothesis. 
However, unmodeled propagation effects can bias the inference of remnant properties and the interpretation of ringdown features. 
We investigate the impact of gravitational lensing on ringdown inference and tests of GR using GW150914-like simulated signals spanning the long-wavelength, wave-optics, and geometric-optics regimes. 
We perform complementary time-domain post-merger and frequency-domain full inspiral--merger--ringdown analyses using unlensed waveform models. 
We find that lensing in the wave-optics regime can produce substantial biases in the inferred remnant mass and spin and can strongly affect the inferred modal content of the ringdown. 
In particular, it can spuriously favor the presence of the first overtone, with $\log_{10}\mathcal{B}^{221}_{220}\sim40$, even though the injected signal has no support for it. 
Parameterized ringdown tests can likewise yield strong apparent support for deviations from the Kerr quasinormal mode (QNM) spectrum, with $\log_{10}\mathcal{B}^{\rm modGR}_{\rm GR}\sim8$. 
In contrast, the evidence for higher-order modes remains largely unchanged for the simulated signals considered here. 
The parameterized analysis using the entire signal further shows that microlensing can bias the inferred frequency of the dominant $(2,2,0)$ QNM. 
These results demonstrate that unmodeled microlensing can produce systematic biases and spurious signatures in ringdown analyses, potentially affecting precision tests of the Kerr hypothesis.
\end{abstract}


\section{Introduction}
\label{sec:intro}

When two black holes (BHs) merge, the highly distorted remnant spacetime relaxes to a Kerr BH by emitting gravitational radiation, in a process often likened to a struck bell ``ringing down''~\citep{Vishveshwara:1970zz,Press:1971wr,Press:1973zz,Teukolsky:1973ha}. 
This final stage of a compact binary coalescence (CBC) is known as the \emph{ringdown}. 
The late-ringdown signal can be modeled as a superposition of quasinormal modes (QNMs)~\citep{Chandrasekhar:1984siy,Berti:2025hly}, which are exponentially damped sinusoids characterized by oscillation frequencies, damping times, and excitation amplitudes. 
According to the no-hair conjecture of GR~\citep{Israel:1967wq,Carter:1971zc,Hawking:1971vc,Ruffini:1971bza}, the QNM spectrum of an astrophysical BH is determined entirely by its mass and spin. 
Consequently, measurements of QNM frequencies and damping times provide a unique probe of the strong-field, highly dynamical regime of gravity. 
In particular, the simultaneous measurement of two or more QNMs enables BH spectroscopy~\citep{Dreyer:2003bv,Berti:2005ys,Berti:2025hly}, providing a test of the Kerr hypothesis and the no-hair conjecture.

Since the first gravitational-wave (GW) detection~\citep{LIGOScientific:2016aoc}, ringdown analyses of CBCs have provided increasingly stringent tests of the remnant properties and the Kerr hypothesis. 
Early studies of GW150914 demonstrated that the dominant $(\ell,m,n)=(2,2,0)$ mode could be identified in the signal and that the remnant mass and spin inferred from the ringdown were consistent with those obtained from full inspiral--merger--ringdown (IMR) analyses~\citep{LIGOScientific:2016lio}. 
Subsequent studies extended the analysis to earlier times and reported evidence for the first overtone $(2,2,1)$~\citep{Isi:2019aib,Finch:2022ynt}, although its interpretation remains sensitive to waveform modeling and the choice of analysis start time~\citep{Cotesta:2022pci,Isi:2022mhy,Isi:2023nif,Carullo:2023gtf}. 
Ringdown analyses of GW190521~\citep{LIGOScientific:2020iuh} further demonstrated the potential for BH spectroscopy through searches for subdominant multipole moments and independent measurements of the remnant properties~\citep{LIGOScientific:2020ufj}. 
More recently, GW240621\_195059 provided evidence for a subdominant ringdown mode consistent with the first overtone, while the high signal-to-noise ratio (SNR) event, GW250114\_082203, enabled stringent constraints on parameterized deviations from the Kerr QNM spectrum~\citep{LIGOScientific:2026wpt,LIGOScientific:2026oim}. 
Despite these increasingly detailed ringdown analyses, the results reported by the LIGO--Virgo--KAGRA (LVK) collaboration remain consistent with the predictions of GR. 
At the same time, the interpretation of subdominant ringdown content can depend sensitively on waveform modeling assumptions, analysis start time, and the statistical treatment of additional modes, underscoring the need to understand potential sources of waveform systematics.

A potential source of waveform systematics in such analyses is gravitational lensing, which can modify the observed GW signal during its propagation from the source to the detector~\citep{Einstein:1936llh,Zwicky:1937zzb}. 
While galaxy-scale lenses produce multiple temporally resolvable images in the geometric-optics regime~\citep{1971NCimB...6..225L,Schneider:1992bmb,Wang:1996as}, compact lenses with masses of approximately $10$--$10^5\,M_\odot$ can produce wave-optics effects (\emph{microlensing}) in the frequency band of ground-based GW detectors, introducing frequency-dependent amplitude and phase modulations in the observed signal~\citep{1986ApJ...307...30D,Nakamura:1997sw,Nakamura:1999uwi,Takahashi:2003ix}. 
Recent studies have shown that these propagation-induced distortions can significantly bias GW parameter estimation (PE)~\citep{Jung:2017flg,Bondarescu:2022srx,Mishra:2023ddt}, mimic orbital eccentricity~\citep{Mishra:2025dpa}, resemble overlapping GW signals~\citep{Rao:2025poe}, and generate apparent deviations from general relativity (GR) in inspiral, propagation, and IMR consistency tests when lensing is neglected in the waveform model~\citep{Mishra:2023vzo,Kopty:2026fnp}.

In this work, we investigate whether microlensing can similarly bias ringdown-based tests of GR, focusing on the recovery of remnant properties, searches for overtones and higher-order modes, and parameterized tests of the Kerr QNM spectrum. 
Using GW150914-like lensed signals and unlensed waveform models for recovery, we quantify how propagation-induced waveform distortions can be absorbed into ringdown parameters or interpreted as additional ringdown content or apparent departures from the Kerr spectrum. 
We perform complementary ringdown analyses with \texttt{pyRing}~\citep{Carullo:2019flw,pyRing} and \texttt{pSEOB}~\citep{Ghosh:2021mrv}, probing the impact of microlensing on QNM inference.
The former provides a time-domain analysis of the post-merger signal, while the latter performs a frequency-domain parameterized analysis of the full-IMR signal. 
As ringdown analyses become increasingly sensitive with future observing runs and next-generation detectors, understanding such astrophysical systematics will be essential for robust precision tests of GR.

The remainder of this paper is organized as follows. 
In Sec.~\ref{sec:theory}, we summarize the basics of GW microlensing and the ringdown modeling. 
Section~\ref{sec:methods} describes the simulated signals (\emph{injections}), inference framework, and analysis methodology. 
We present our results in Sec.~\ref{sec:results} and summarize our conclusions and their implications for future ringdown tests of GR in Sec.~\ref{sec:conclude}.

\section{Microlensing and Ringdown Formalism}\label{sec:theory}

\subsection{Microlensing of Gravitational Waves}
\label{subsec:microlensing}

GWs propagating through an inhomogeneous gravitational potential can be gravitationally lensed by intervening matter. 
In the geometric-optics regime, lensing produces multiple images with different magnifications, time delays, and Morse phases~\citep{Schneider:1992bmb}, while in the wave-optics regime diffraction becomes important and the individual images cannot be treated independently~\citep{Nakamura:1997sw,Takahashi:2003ix}. 
For compact lenses whose characteristic scale is comparable to the GW wavelength, lensing therefore produces frequency-dependent amplitude and phase modulations in the observed waveform. 

In the wave-optics regime, the lensed waveform can be written in the frequency domain as~\citep{Nakamura:1997sw,Takahashi:2003ix}
\begin{equation}
    \tilde h_{\rm L}(f) = F(f;\boldsymbol{\Lambda})\,\tilde h_{\rm U}(f;\boldsymbol{\theta}),
    \label{eq:lensed-waveform}
\end{equation}
where $\tilde h_{\rm L}$ and $\tilde h_{\rm U}$ denote the lensed and unlensed frequency-domain waveforms, respectively, $\boldsymbol{\theta}$ denotes the usual binary parameters, and $\boldsymbol{\Lambda}$ denotes the lens parameters. 
For an isolated point-mass lens, we parameterize the lens by the redshifted lens mass $M_{\rm Lz}=(1+z_{\rm L})M_{\rm L}$, where $M_{\rm L}$ and $z_{\rm L}$ are the lens mass and redshift, respectively, and by the dimensionless source position $y$, defined as the source--lens separation in units of the Einstein radius. 
The corresponding amplification factor has an analytic solution,
\begin{align}
    F(w,y) ={}&\exp\!\left[\frac{\pi w}{4} +\frac{iw}{2}
        \left(\ln\frac{w}{2}-2\phi_m(y)\right)\right]\nonumber\\
        &\times \Gamma\!\left(1-\frac{iw}{2}\right)\,
        {}_1F_1\!\left(\frac{iw}{2},1;\frac{iw y^2}{2}\right),
        \label{eq:F_pointmass}
\end{align}
where $w = \frac{8\pi G M_{\rm Lz} f}{c^3}$ is the dimensionless frequency and $\phi_m(y)$ is the Fermat potential evaluated at the minimum of the time-delay surface. 

The transition between the wave- and geometric-optics regimes is controlled by the dimensionless frequency $w$. 
For $w\ll1$, corresponding to GW wavelengths much larger than the characteristic lens scale, the amplification factor approaches unity and the lensing effect is strongly suppressed. 
For $w\sim1$, diffraction produces pronounced frequency-dependent modulations, and in the high-frequency limit $w\gg1$, the amplification factor approaches the geometric-optics form
\begin{equation}
    F(f) \simeq\sum_j\sqrt{|\mu_j|}\exp\!\left[
        2\pi i f t_{d,j}-i\pi n_j\right],
    \label{eq:F_geo}
\end{equation}
where $\mu_j$, $t_{d,j}$, and $n_j$ are the magnification, time delay, and Morse index\citep{Schneider:1992bmb,Dai:2017huk} of the $j$-th image, respectively. 

For convenience, we characterize the lensing regime using the frequency $f_{\rm ML}$ defined by $w(f_{\rm ML})=1$~\citep{Mishra:2023ddt},
\begin{equation}
    f_{\rm ML}\equiv\frac{c^3}{8\pi G M_{\rm Lz}}
        \simeq0.8\times10^3\,{\rm Hz}\left(\frac{10\,M_\odot}{M_{\rm Lz}}\right).
    \label{eq:fML}
\end{equation}
Thus, for a fixed lens mass, frequencies $f\ll f_{\rm ML}$ correspond to the long-wavelength limit, frequencies $f\sim f_{\rm ML}$ probe the wave-optics regime, and frequencies $f\gg f_{\rm ML}$ approach the geometric-optics limit. 
This classification is useful for understanding the lens parameter space, although the observational impact of lensing also depends on the source, the impact parameter, and the amount of signal power present at the frequencies where the lensing modulations occur.
 
Searches for microlensing in GW observations have so far found no confirmed detections~\citep{Hannuksela:2019kle,LIGOScientific:2021izm,LIGOScientific:2023bwz,Janquart:2023mvf,LIGOScientific:2025cwb}. 
However, GW231123\_135430 was identified as a notable outlier in the microlensing analysis, although the evidence for it remains inconclusive~\citep{LIGOScientific:2025cwb}.
Note that we do not model complex lens environments or composite lens systems, such as microlenses embedded in galaxy-scale potentials~\citep{Diego:2019lcd,Mishra:2021xzz} in this work. 
Such configurations can combine wave-optics distortions with the magnification and time-delay structure of a macroscopic lens and require a more general treatment.

\subsection{Ringdown and Quasinormal Modes}
\label{subsec:ringdown}

According to GR, the remnant formed following the merger of a binary BH (BBH) settles toward a stationary Kerr BH through the emission of gravitational radiation. 
In the late post-merger phase, when the remnant spacetime can be treated as a perturbation of the Kerr geometry, the emitted radiation is described as a superposition of QNMs~\citep{Vishveshwara:1970cc,Teukolsky:1973ha,Chandrasekhar:1984siy}. 
They are exponentially damped oscillations characterized by complex frequencies. 
For an astrophysical BH with negligible charge, the QNM frequencies and damping times are determined by the remnant mass $M_{\rm f}$ and dimensionless spin $\chi_{\rm f}$, as expected from the no-hair conjecture~\citep{Carter:1971zc,Hawking:1971vc,Robinson:1975bv}. 
Measurements of multiple QNMs therefore provide a probe of the remnant and enable BH spectroscopy, offering a direct test of the Kerr hypothesis~\citep{Dreyer:2003bv,Berti:2009kk,Berti:2025hly}.

The ringdown waveform can be expressed schematically as
\begin{align}
    h_+(t)-i &h_\times(t)
    = \sum_{\ell=2}^{+\infty}\sum_{m=-\ell}^{\ell}\sum_{n=0}^{+\infty} 
    \mathcal{A}_{\ell mn}\exp\left[-\frac{t-t_0}{\tau_{\ell mn}}\right] \nonumber \\
    & \exp\left[i\omega_{\ell mn}(t-t_0)\right]
    {}_{-2}S_{\ell mn}(\iota,\varphi;\chi_{\rm f})\,,
    \label{eq:ringdown-waveform}
\end{align}
where $(\ell,m,n)$ denote the angular and overtone indices, $\mathcal{A}_{\ell mn}$ is the complex excitation amplitude, $\omega_{\ell mn}=2\pi f_{\ell mn}$ is the oscillation frequency, $\tau_{\ell mn}$ is the damping time, and ${}_{-2}S_{\ell mn}$ are spin-weighted spheroidal harmonics. 
For a given $(\ell,m)$, the fundamental mode ($n=0$) is the least damped and dominates the signal at sufficiently late times, whereas overtones ($n>0$) decay more rapidly and can contribute at earlier times after merger. 
Higher-order modes can also contribute to the observed ringdown, depending on their excitation amplitudes and the properties of the progenitor binary.

Ringdown tests of GR exploit these properties in several complementary ways. 
The remnant parameters inferred from the ringdown can be compared with those inferred from the full IMR signal, while measurements of multiple QNMs can be used to test the consistency of the observed spectrum with that predicted for a Kerr BH~\citep{Brito:2018rfr,Carullo:2019flw,Isi:2019aib}.
Alternatively, the QNM frequencies and damping times can be allowed to deviate phenomenologically from their GR values, providing theory-independent tests of the Kerr spectrum~\citep{Carullo:2019flw,Isi:2019aib,Ghosh:2021mrv,Ma:2022wpv}.
These approaches form part of the standard suite of ringdown-based tests of GR routinely performed by the LVK collaboration~\citep{LIGOScientific:2019fpa,LIGOScientific:2020tif,LIGOScientific:2021sio,LIGOScientific:2026wpt,LIGOScientific:2026oim}.
The details of these analyses are given in Sec.~\ref{subsec:ringdown_tests}.

\section{GW Data Analysis Framework}\label{sec:methods}

\subsection{Fitting-factor study}
\label{subsec:FF}

As an initial diagnostic of the impact of microlensing on the ringdown waveform, we compute the fitting factor (FF) between the lensed injections and the unlensed waveform templates. 
The FF provides a computationally inexpensive measure of the extent to which the microlensing-induced waveform distortions can be absorbed by changes in the parameters of an unlensed binary model, without requiring a full Bayesian PE analysis.

The overlap between two waveforms $h_1$ and $h_2$ is defined as
\begin{equation}
    \mathcal{O}[h_1,h_2] = \frac{\langle h_1,h_2\rangle}
            {\sqrt{\langle h_1,h_1\rangle\langle h_2,h_2\rangle}},
    \label{eq:overlap}
\end{equation}
where $\langle\cdot,\cdot\rangle$ denotes the noise-weighted inner product. 
The match, $\mathcal{M}$, is obtained by maximizing the overlap over the coalescence time and phase. 
The FF is then defined as the maximum match over the intrinsic parameters of the template family~\citep{Apostolatos:1995pj},
\begin{equation}
    \mathrm{FF} = \max_{\{\lambda\}}\mathcal{M}(\lambda)
                = \frac{\rho_{\rm rec}}{\rho_{\rm opt}},
    \label{eq:FF}
\end{equation}
where $\lambda$ denotes the parameters over which the maximization is performed, $\rho_{\rm opt}$ is the optimal SNR of the signal, and $\rho_{\rm rec}$ is the SNR recovered by the best-fit template.
The corresponding mismatch, $1-\mathrm{FF}$, measures the fractional loss in recovered SNR due to the inability of the template family to reproduce the signal morphology~\citep{Ajith:2012mn}.

We compute the FF for GW150914-like lensed injections over a two-dimensional lens-parameter space spanned by the dimensionless impact parameter $y$ and the redshifted lens mass $M_{\rm Lz}$. 
We consider $y\in[10^{-2},2]$, sampled logarithmically at $40$ values, and $M_{\rm Lz}\in[10,10^5]\,M_\odot$, sampled logarithmically at $50$ values. 
The parameter ranges are chosen to also include weakly lensed configurations while avoiding a parameter region where the lensing-induced waveform distortions are negligible for the frequency band considered here.
The resulting $2000$ lens configurations span the long-wavelength, wave-optics, and geometric-optics regimes relevant to the ground-based GW frequency band.

\begin{figure}[htbp] \centering \includegraphics[width=\linewidth]{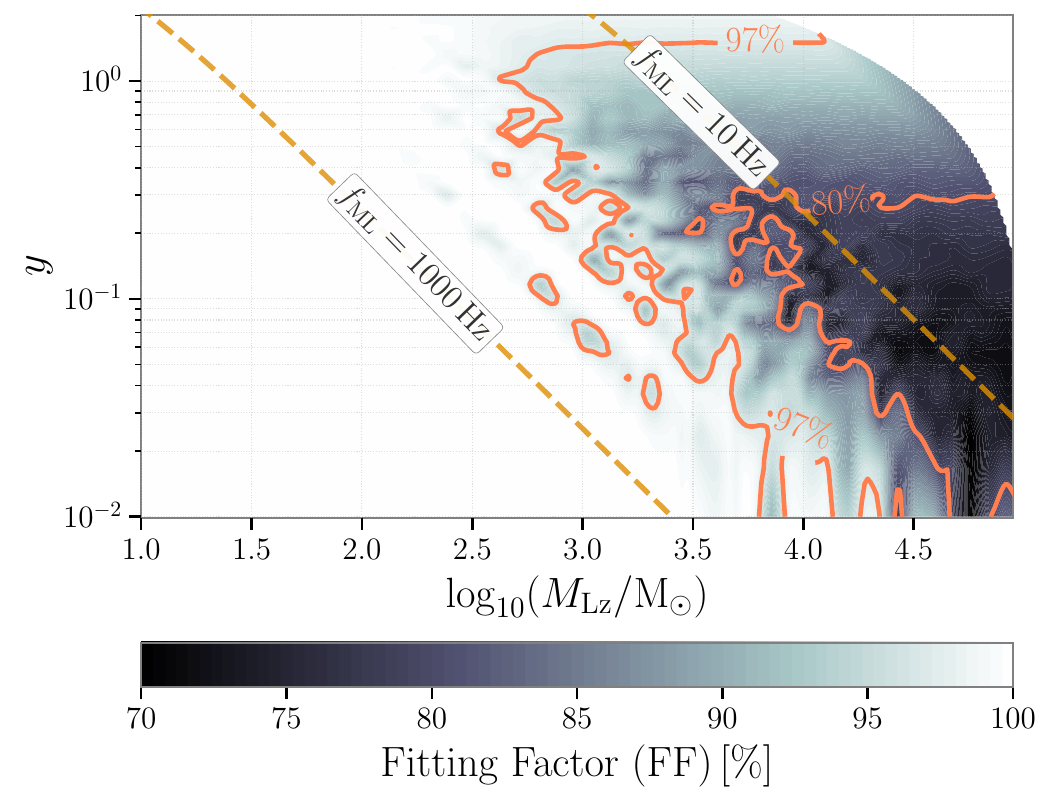} 
\caption{
Fitting factors for GW150914-like lensed injections as a function of impact parameter $y$ and redshifted lens mass $M_{\rm Lz}$. 
Solid orange contours indicate FFs of 97\% and 80\%. 
The dashed lines, corresponding to $f_{\rm ML}=1000$ Hz and $10$ Hz, indicate characteristic frequencies used to organize the lens-parameter space into long-wavelength, wave-optics, and geometric-optics regimes. 
A significant reduction in the fitting factor is observed in the wave-optics and geometric-optics regimes, indicating pronounced waveform distortions relative to unlensed GR templates. 
The region in the top-right corner is excluded because it corresponds to cases where the time delay between the two microimages exceeds the chirp duration, i.e., cases exhibiting resolved lensed images.
} 
\label{fig:FF_result} 
\end{figure} 

The unlensed binary waveform is generated using the \texttt{NRSur7dq4}~\citep{Varma:2019csw} waveform approximant, with the intrinsic parameters fixed to the median values of the publicly available GW150914 posterior samples~\citep{LIGOScientific:2022GWTC2POINT1PE}\footnote{We use the \texttt{\scriptsize{C01:Mixed}} samples from ~\\ \texttt{\scriptsize{IGWN-GWTC2p1-v2-GW150914\_095045\_PEDataRelease\_mixed\_cosmo.h5}}.}. 
The lensing amplification factor, given by Eq.~\eqref{eq:F_pointmass}, is then applied in the frequency domain to obtain the corresponding lensed signals, using the publicly available package \textsc{gwmat}\footnote{\url{https://git.ligo.org/anuj.mishra/gwmat}}~\citep{Mishra:2023ddt}. 
We use the Advanced LIGO power spectral density (PSD) at the design sensitivity in evaluating the noise-weighted overlaps~\citep{KAGRA:2013rdx}\footnote{\url{https://dcc.ligo.org/LIGO-P1200087-v19/public}}
Since this study is intended to characterize the effect of microlensing on the ringdown part of the signal, the overlap is evaluated from the fundamental QNM frequency $f_{220}$\footnote{We estimate $f_{220}$ using the fitting coefficients for $M_{\rm f}\omega_{220}$ given in Table VIII of~\citet{Berti:2005ys}.} up to the Nyquist frequency of $1024\,{\rm Hz}$.

For each lens configuration, the FF is obtained by maximizing the match over the four-dimensional intrinsic parameter space of the unlensed template, consisting of the chirp mass $\mathcal{M}$, mass ratio $q$, and the two aligned spin components $\chi_{\rm 1z}$ and $\chi_{\rm 2z}$. 
We restrict the maximization to aligned spins $(\chi_{\rm 1z}, \chi_{\rm 2z})$ to ensure reliable convergence, which is difficult to achieve when varying all six spin components. 
To isolate the mismatch arising from unmodeled lensing rather than precession bias, the in-plane spin components of the template are held fixed at their injected values throughout the maximization.
We perform this maximization using the Nelder--Mead algorithm~\citep{Nelder:1965zz}, as implemented in the \texttt{scipy.optimize} module~\citep{Virtanen:2019joe}. 

Figure~\ref{fig:FF_result} shows the resulting FF across the lens parameter space. 
The dashed curves correspond to $f_{\rm ML}=10$ and $1000\,{\rm Hz}$ and indicate the characteristic boundaries used to distinguish the long-wavelength, wave-optics, and geometric-optics regimes. 
The FF remains close to unity in the long-wavelength regime, where the lensing-induced modification of the waveform is weak. 
It decreases substantially in the wave-optics regime, reaching values as low as $\sim0.7$, indicating that the frequency-dependent amplitude and phase modulations cannot be efficiently reproduced by any choice of parameters in the unlensed waveform family. 
The region in the upper-right corner of Fig.~\ref{fig:FF_result} is excluded from the analysis because the time delay between the microimages becomes longer than the chirp duration. 
Such configurations correspond to resolvable multiple lensed signals and therefore fall outside the single distorted lensed signal scenario considered here.

It is important to distinguish a low FF from a bias in any particular inferred parameter. 
The FF measures the overall distinguishability of the microlensed signal from the unlensed waveform family, whereas parameter biases depend on the projection of the waveform distortion onto individual directions in the parameter space.
Consequently, a substantial mismatch does not by itself imply a correspondingly large bias in the QNM frequencies, damping times, or mode amplitudes. 
This distinction is particularly relevant in the geometric-optics regime, where the lensed waveform approaches a superposition of separate microimages carrying the same intrinsic QNM spectrum.
Whether the resulting mismatch translates into a significant bias in ringdown inference therefore requires an explicit Bayesian analysis, which we perform in the following sections.

\subsection{Ringdown Tests of GR}
\label{subsec:ringdown_tests}

\paragraph{\textbf{Simulated signals:}}
We generate GW150914-like lensed injections using the quasi-circular, non-precessing numerical relativity (NR) simulation \texttt{SXS:BBH:0318} from the SXS catalog~\citep{Boyle:2019kee}, with mass ratio $q=1.22$ and dimensionless aligned spin components $\chi_{\rm 1z}=0.33$ and $\chi_{\rm 2z}=-0.44$.
The total mass and extrinsic parameters are fixed to the median values of the GW150914 posterior samples~\citep{LIGOScientific:2022GWTC2POINT1PE}, as in the FF analysis.
The zero-noise injections are analyzed in a three-detector network consisting of the LIGO Hanford (H), LIGO Livingston (L) and Virgo (V) detectors operating at their projected Advanced LIGO~\citep{LIGOScientific:2014pky} and Advanced Virgo~\citep{VIRGO:2014yos} design sensitivities~\citep{KAGRA:2013rdx}.

For the ringdown analyses, the luminosity distance is adjusted such that the resulting post-merger signal has an optimal network SNR of $15$. 
We define the reference post-merger start time $t_0$ as the time of the maximum of the strain amplitude squared ($h_+^2(t) + h_\times^2(t)$).
Since the peak time can itself be affected by waveform distortions, we determine $t_0$ separately for each injection using an initial unlensed GR PE analysis. 
Specifically, we perform full-IMR Bayesian PE with \texttt{Bilby}~\citep{Ashton:2018jfp,Smith:2019ucc}, using the
\texttt{IMRPhenomXPHM-SpinTaylor} waveform model~\cite{Khan:2018fmp, Garcia-Quiros:2020qpx, Pratten:2020ceb, Colleoni:2024knd}. 
From a randomly selected subset of the resulting posterior samples, we reconstruct the two GW polarizations and project them onto the Hanford detector, which exhibited the highest SNR among the HLV detectors considered. 
We then compute the time of the maximum of $h_+^2 + h_\times^2$ for each sample to obtain a marginalized posterior for $t_0$. 
The median of this peak strain distribution is adopted as the reference peak time $t_0$ for each injection. 

We sample the lensing parameter space on a logarithmic grid of $10$ values in each of the redshifted lens mass and dimensionless impact parameter, covering $M_{\rm Lz}\in[10,10^{5}]\,M_\odot$ and  $y\in[10^{-2},1]$, for a total of $100$ initial lens configurations. 
We exclude configurations for which the time delay between the lensed images exceeds the duration of the simulated signal, since such systems would produce temporally resolvable images rather than the single lensed signal considered here. 
After this exclusion, the final sample contains $96$ injections used in the ringdown analyses.

\paragraph{\textbf{Ringdown analysis with \texttt{pyRing}:}}
We perform a time-domain Bayesian analysis of the post-merger signal using \texttt{pyRing}~\citep{Carullo:2019flw,pyRing},   where the ringdown is modeled as a superposition of damped sinusoids whose frequencies and damping times are determined by the mass and spin of the remnant Kerr BH. 
For each microlensed injection, we analyze using three waveform models that differ in the physical content included in the ringdown:
\begin{itemize}
    \item $\mathrm{Kerr}_{220}$, containing only the fundamental $(\ell,m,n)=(2,2,0)$ mode;
    \item $\mathrm{Kerr}_{221}$, containing the $(2,2,0)$ fundamental mode and the first overtone $(2,2,1)$;
    \item $\mathrm{Kerr}_{\mathrm{HM}}$, containing the fundamental mode and higher multipoles with $\ell\leq4$ ( and $n=0$).
\end{itemize}
These models allow us to separately assess whether lensing can be absorbed into the dominant QNM, spuriously favor the presence of an overtone, or produce apparent evidence for higher-order modes.

The choice of ringdown start time is model dependent. 
For the $\mathrm{Kerr}_{221}$ analysis, the ringdown model starts at the reference time $t_0$, allowing the more rapidly damped overtone to contribute at earlier times. 
For the $\mathrm{Kerr}_{220}$ and $\mathrm{Kerr}_{\mathrm{HM}}$ analyses, we instead start at $t_0+10GM_{\rm f}/c^3$, where the contribution from the least-damped fundamental mode is expected to dominate. 
We adopt uniform priors on the remnant mass, spin, and mode amplitudes. 
The sky-location and polarization parameters $(\alpha,\delta,\psi)$ are fixed to their median values obtained from the corresponding IMR analysis, while the inclination and azimuthal angles are fixed to $(\iota,\phi)=(\pi,0)$. 
This choice reduces degeneracies between the mode amplitudes and phases, particularly in analyses including overtones~\citep{Cotesta:2022pci}.

We quantify the statistical preference for additional QNM content through Bayes factors between competing ringown models. 
In particular,
\begin{equation}
    \mathcal{B}^{221}_{220}
    =
    \frac{\mathcal{Z}_{221}}{\mathcal{Z}_{220}}
\end{equation}
quantifies the preference for including the $(2,2,1)$ overtone, while
\begin{equation}
    \mathcal{B}^{\mathrm{HM}}_{220}
    =
    \frac{\mathcal{Z}_{\mathrm{HM}}}{\mathcal{Z}_{220}}
\end{equation}
quantifies the preference for subdominant multipoles relative to the dominant-mode model. 
Here, $\mathcal{Z}$ denotes the Bayesian evidence for the corresponding ringdown model. 

We additionally perform a parameterized test of the Kerr QNM spectrum using the $\mathrm{Kerr}_{221}$ model. 
We allow the frequency and damping time of the $(2,2,1)$ mode to deviate fractionally from their GR predictions,
\begin{equation}
    f_{221}
    =
    f_{221}^{\rm GR}
    \left(1+\delta\hat f_{221}\right),
    \qquad
    \tau_{221}
    =
    \tau_{221}^{\rm GR}
    \left(1+\delta\hat\tau_{221}\right),
\end{equation}
while keeping the fundamental $(2,2,0)$ mode fixed to its GR-predicted frequency and damping time. 
We adopt uniform priors on $\delta\hat f_{221}\in[-1,1]$ and $\delta\hat\tau_{221}\in[-0.9,1]$. 
The GR hypothesis corresponds to $\delta\hat f_{221}=\delta\hat\tau_{221}=0$.

Following the model-selection prescription used in parameterized ringdown tests~\citep{LIGOScientific:2020tif,LIGOScientific:2021sio}, we consider three modified-GR sub-hypotheses: a deviation in the $(2,2,1)$ frequency only, a deviation in its damping time only, and simultaneous deviations in both quantities. 
With equal prior weights assigned to the three sub-hypotheses, we define the combined Bayes factor as
\begin{equation}
    \mathcal{B}^{\mathrm{modGR}}_{\mathrm{GR}}
    =
    \frac{1}{3}
    \left(
    \mathcal{B}^{\delta f_{221}}_{\mathrm{GR}}
    +
    \mathcal{B}^{\delta\tau_{221}}_{\mathrm{GR}}
    +
    \mathcal{B}^{\delta f_{221},\delta\tau_{221}}_{\mathrm{GR}}
    \right).
    \label{eq:logB_modGR_vs_GR}
\end{equation}
A value of $\mathcal{B}^{\mathrm{modGR}}_{\mathrm{GR}}>1$ indicates increased support for the modified-GR hypothesis relative to GR.

\paragraph{\textbf{Full-IMR analysis with \texttt{pSEOB}:}}
As a complementary probe, we measure QNM spectrum within the framework of a full IMR waveform model, \texttt{pSEOBNRv4HM}~\citep{Ghosh:2021mrv}. 
This model is a parameterized extension of the spinning effective-one-body waveform, \texttt{SEOBNRv4HM}, including higher modes and calibrated against non-precessing NR simulations~\citep{Cotesta:2018fcv,Capano:2021etf}. 
Unlike the \texttt{pyRing} analysis, this frequency-domain analysis makes use of the total SNR since it models the entire signal with the ringdown start-time built into the model based on NR calibrations.

Within this framework, we introduce fractional deviations in the frequencies and damping times of the fundamental QNM according to
\begin{equation}
\begin{aligned}
    f_{lm0} &=
    f^{\mathrm{GR}}_{\ell m0}
    \left(1+\delta\hat f_{\ell m0}\right),\\
    \tau_{lm0} &=
    \tau^{\mathrm{GR}}_{\ell m0}
    \left(1+\delta\hat\tau_{\ell m0}\right),
    \label{eq:pseob_frac_deviations}
\end{aligned}
\end{equation}
where $f_{\ell m0}$ and $\tau_{\ell m0}$ denote the frequency and damping time of the $(\ell,m,0)$ QNM, respectively, and the GR prediction corresponds to $\delta\hat f_{\ell m0}=\delta\hat\tau_{\ell m0}=0$. 
The parameterized deviations are constrained jointly with the binary parameters using the full signal. 
We use the same set of microlensed injections as in the post-merger analysis, with the signals rescaled to an optimal network SNR of $30$ to ensure enough cycles across the inspiral, merger, and ringdown phases.

\section{Results and Discussions}\label{sec:results}

\subsection{Biases in parameter inference using \texttt{pyRing}}\label{subsec:pyring_results}

\begin{figure*}[htbp]
\centering
\begin{minipage}{0.48\textwidth}
    \centering
    \includegraphics[width=\linewidth]{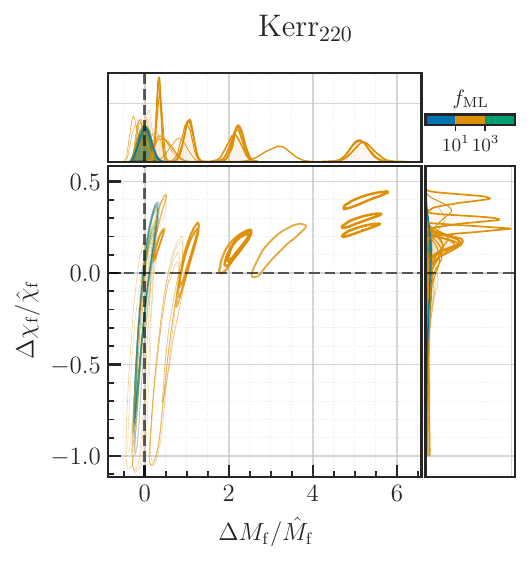}
\end{minipage}
\hspace{0.01\textwidth}
\begin{minipage}{0.48\textwidth}
    \centering
    \includegraphics[width=\linewidth]{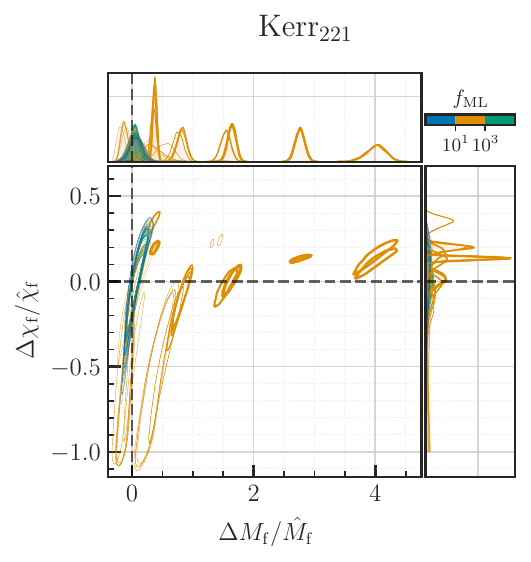}
\end{minipage}
\caption{
Joint 2D posterior distributions showing the $90\%$ credible regions for fractional deviations in the final mass ($M_{\rm f}$) and spin ($\chi_{\rm f}$) recovered using (a) Kerr$_{220}$ and (b) Kerr$_{221}$ ringdown templates, relative to the corresponding values from the full IMR analysis, shown for all $96$ microlensed injections. 
Colors indicate the microlensing frequency parameter $f_{\rm ML}$. 
Significant biases arise predominantly in the wave-optics-dominated regime. 
The linewidths and transparency are scaled according to the statistical significance of the deviation from zero, thereby downweighting broad, uninformative posteriors.
}
\label{fig:rem_prms_combined}
\end{figure*}

We begin by comparing the remnant mass and spin inferred from the ringdown analyses with the corresponding quantities obtained from the full IMR analysis. 
This comparison enables us to quantify systematic biases introduced by microlensing in standard ringdown tests. 
We analyze the results obtained using the three waveform models $\mathrm{Kerr}_{220}$, $\mathrm{Kerr}_{221}$, and $\mathrm{Kerr}_{\rm HM}$. 
We focus on the recovery of the remnant mass and spin using the $\mathrm{Kerr}_{220}$ and $\mathrm{Kerr}_{221}$ models, and subsequently examine the evidence for higher modes using $\mathrm{Kerr}_{\rm HM}$.

Fig.~\ref{fig:rem_prms_combined} shows the joint posterior distributions for the relative deviations in the final mass and spin, $(\Delta M_\mathrm{f}/M_\mathrm{f},\;\Delta\chi_\mathrm{f}/\chi_\mathrm{f})$, recovered using the $\mathrm{Kerr}_{220}$ and $\mathrm{Kerr}_{221}$ templates relative to the IMR analysis.
The colors indicate the microlensing frequency parameter $f_{\mathrm{ML}}$. 
For a signal fully consistent with GR and unaffected by propagation effects, the remnant parameters inferred from the ringdown and IMR analyses are expected to agree, resulting in posteriors centered at $(0,0)$. 
In both the $\mathrm{Kerr}_{220}$ and $\mathrm{Kerr}_{221}$ cases, this behavior is observed for injections in both the long-wavelength and geometric-optics regimes (green and blue contours), where the $90\%$ credible regions consistently enclose the $(0,0)$ point, indicating no statistically significant bias in the recovered final mass and spin. 
In contrast, injections in the wave-optics regime (orange contours) exhibit pronounced deviations. 
While a small subset of these injections remains consistent with GR at the $90\%$ level, the majority show substantial offsets from $(0,0)$, indicating strong biases in the recovered remnant parameters. 
In extreme cases, we observe fractional mass biases $\Delta M_\mathrm{f}/M_\mathrm{f} \gtrsim 1$, reaching values as large as $\sim5$, accompanied by spin biases up to $\Delta\chi_\mathrm{f}/\chi_\mathrm{f} \sim 0.4$. 
These results show that microlensing can produce substantial biases in the inferred remnant properties in the wave-optics regime. 

\paragraph{\textbf{Preference for higher overtones, not higher modes:}}

\begin{figure}[htbp]
\centering
\includegraphics[width=\columnwidth]{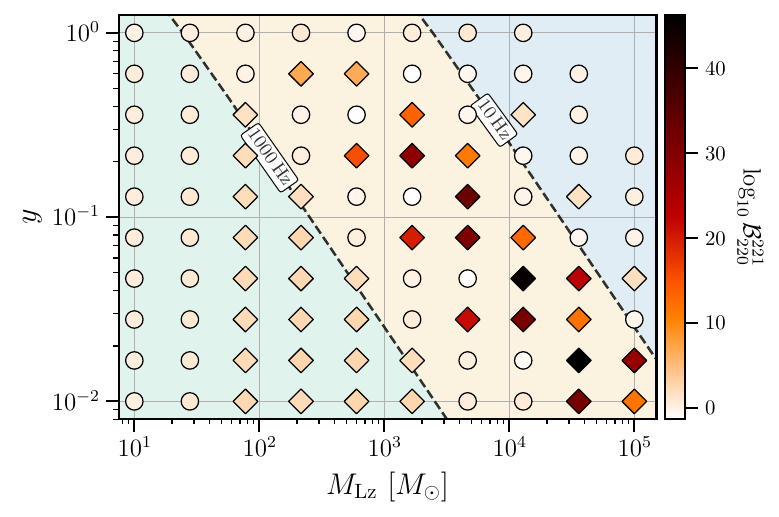}
\vspace{0.5em}
\includegraphics[width=\columnwidth]{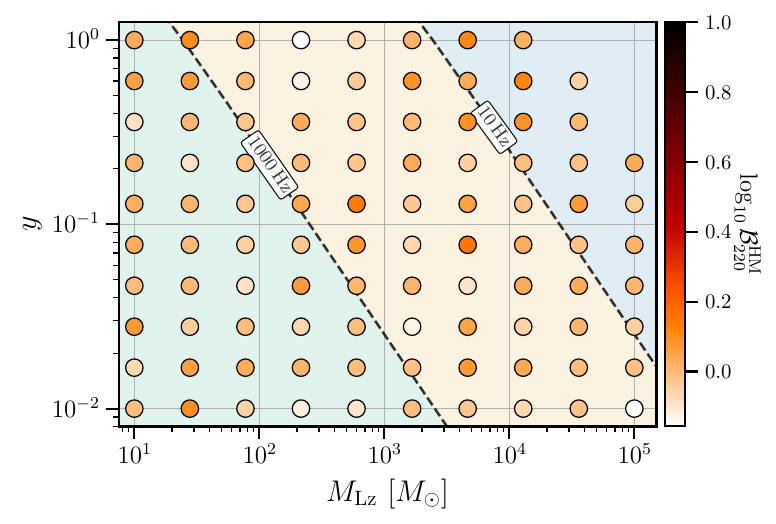}
\caption{
Logarithmic Bayes factors for GW150914-like microlensed injections as a function of the lens parameters. 
The top and bottom panels show, respectively, the overtone Bayes factor $\log_{10}\mathcal{B}^{221}_{220}$ and the higher-mode Bayes factor $\log_{10}\mathcal{B}^{\mathrm{HM}}_{220}$, with $\mathrm{Kerr}_{220}$ as the reference model. 
The black dashed contours correspond to $f_{\mathrm{ML}}=10~\mathrm{Hz}$ and $f_{\mathrm{ML}}=10^3~\mathrm{Hz}$, approximately delineating the wave-optics regime. 
The preference for the overtone model is strongly enhanced for a subset of wave-optics microlensed signals, whereas the higher-mode Bayes factor remains close to that of the unlensed reference case across the lens parameter space.
}
\label{fig:Bayes_factor_combined}
\end{figure}

We next investigate whether microlensing-induced distortions can be absorbed by additional QNM degrees of freedom, thereby producing spurious evidence for subdominant ringdown modes. 
The top panel of Fig.~\ref{fig:Bayes_factor_combined} shows the Bayes factor $\log_{10}\mathcal{B}^{221}_{220}$ for including the first overtone. 
For the unlensed reference injection, analyzed using the same NR waveform and binary source parameters as the lensed injections, we obtain $\log_{10}\mathcal{B}^{221}_{220}=-0.88$, indicating no preference for the overtone model. 
The lensed injections exhibit a systematic departure from this behavior.
The Bayes factor increases toward the wave-optics regime of the lens parameter space, where interference effects are the strongest, reaching values as high as $\log_{10}\mathcal{B}^{221}_{220}\sim40$. 
The clustering of high Bayes factor injections in this region indicates that microlensing can significantly distort the ringdown morphology in a manner that mimics the presence of overtones.

The bottom panel of Fig.~\ref{fig:Bayes_factor_combined} shows the corresponding Bayes factor $\log_{10}\mathcal{B}^{\mathrm{HM}}_{220}$ for the inclusion of subdominant multipoles. 
For the unlensed reference injection, we find $\log_{10}\mathcal{B}^{\mathrm{HM}}_{220}=0.07$, consistent with the absence of measurable higher-mode content expected for the GW150914-like simulated signal considered here. 
In contrast to the overtone analysis, the lensed injections do not produce any appreciable biases in the higher-mode Bayes factor. 
Across the full lens parameter space, $\log_{10}\mathcal{B}^{\mathrm{HM}}_{220}$ remains within the narrow range $[-0.15,0.15]$, including in the wave-optics regime.
For the configurations considered here, microlensing therefore produces a pronounced preference for the overtone model without a comparable enhancement of the higher-mode model.

\paragraph{\textbf{Apparent deviations from the Kerr spectrum:}}

\begin{figure}[htbp]
    \centering
    \includegraphics[width=\linewidth]{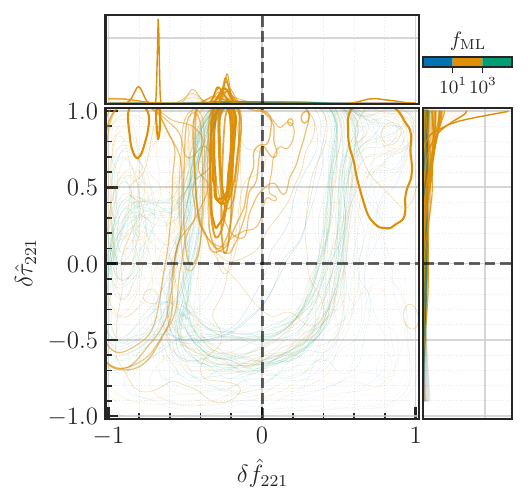}
    \caption{
    Joint $90\%$ credible regions for the fractional deviations in the frequency, $\delta\hat{f}_{221}$, and damping time, $\delta\hat{\tau}_{221}$, of the $(2,2,1)$ mode obtained with the \texttt{pyRing} analysis. 
    The GR prediction corresponds to $(\delta\hat{f}_{221},\delta\hat{\tau}_{221})=(0,0)$. 
    Significant frequency deviations occur predominantly for lensed signals in the wave-optics regime, whereas the damping-time deviations are generally weakly constrained.
    }
    \label{fig:df_dt_221}
\end{figure}

Figure~\ref{fig:df_dt_221} shows the $90\%$ joint credible regions for the fractional deviations in the frequency and damping time of the $(2,2,1)$ mode, obtained from the \texttt{pyRing} analysis allowing both $\delta\hat{f}_{221}$ and $\delta\hat{\tau}_{221}$ to vary simultaneously.
The inferred frequency deviation $\delta\hat{f}_{221}$ departs from the GR value for a subset of the lensed
injections, with the most pronounced deviations occurring in the wave-optics regime. 
In contrast, the damping-time deviation $\delta\hat{\tau}_{221}$ remains largely unconstrained, with broad posteriors often railing against the prior boundaries. 
We therefore do not interpret the damping-time posterior as providing evidence for a physical deviation from the GR prediction and focus primarily on the frequency deviation in what follows.

\begin{figure}[htbp]
    \centering
    \includegraphics[width=\linewidth]{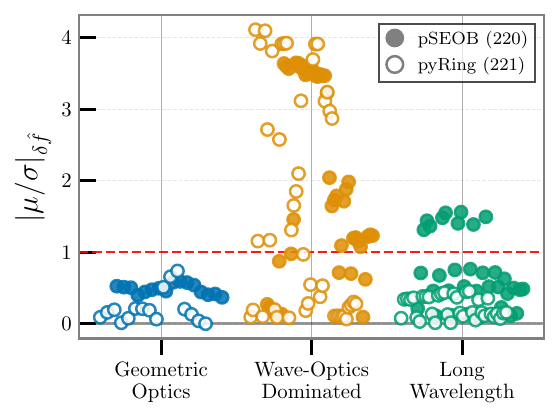}
    \caption{
    Absolute statistical significance, $\left|\mu/\sigma\right|_{\delta\hat{f}}$, of the fractional frequency deviation across the three lensing regimes, ordered by increasing $f_{\rm ML}$. 
    Filled and open circles denote the \texttt{pSEOB} (220) and \texttt{pyRing} (221) recoveries, respectively. 
    The horizontal dashed line marks unit statistical significance, $|\mu/\sigma|=1$, while the solid line indicates the
    GR prediction, $\delta\hat{f}=0$. 
    Significant frequency biases are concentrated in the wave-optics regime. 
    In the long-wavelength regime, several tightly constrained posteriors exceed the unit significance threshold, but their absolute deviations remain small, with $|\delta\hat{f}|\sim0.03$.
    }
    \label{fig:df_221_220}
\end{figure}

The statistical significance of the inferred frequency deviations is shown more explicitly in Fig.~\ref{fig:df_221_220}, where we plot $\left|\mu/\sigma\right|_{\delta\hat f}$ as a function of the characteristic microlensing frequency $f_{\rm ML}$. 
The open circles show the \texttt{pyRing} results for the $(2,2,1)$ mode, while the filled circles show the corresponding \texttt{pSEOB} results for the $(2,2,0)$ mode discussed in Sec.~\ref{subsec:pseob_results}.
The largest frequency biases occur for signals in the wave-optics regime, where the microlensing-induced distortions overlap strongly with the frequency range to which the ringdown analysis is sensitive.
A few long-wavelength injections also yield $|\mu/\sigma|_{\delta\hat f}>1$; however, their inferred physical deviations are small, with $|\delta\hat f|\sim0.03$. 
This highlights an important distinction between statistical and physical significance; at sufficiently high SNR, even a small systematic offset can be statistically distinguishable from zero, without implying a physically meaningful departure from the Kerr prediction. 
By contrast, the wave-optics cases can produce both statistically significant and substantially larger frequency biases, demonstrating that the apparent deviations are not merely a consequence of increased statistical precision but arise from the stronger waveform distortions caused by microlensing.

\begin{figure}[htp]
    \centering
    \includegraphics[width=0.48\textwidth]{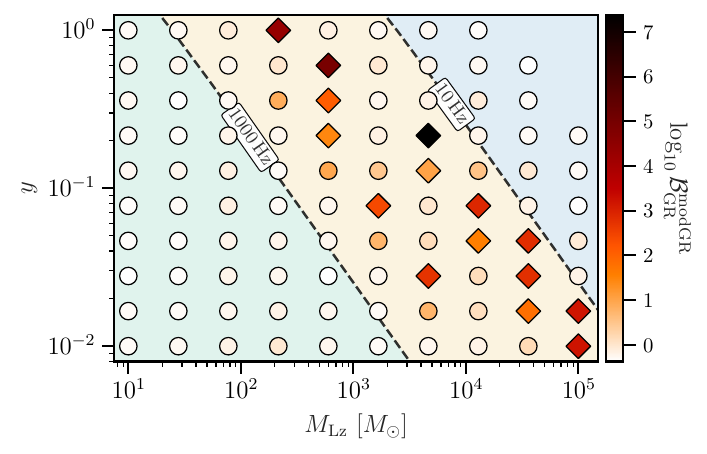}
    \caption{
    Logarithmic Bayes factor $\log_{10}\mathcal{B}^{\rm modGR}_{\rm GR}$ for GW150914-like lensed injections as a function of the lens parameters. 
    The contours are the same as those in Fig.~\ref{fig:Bayes_factor_combined}. 
    A strong preference for the modified-GR hypothesis is obtained for a subset of signals in the wave-optics regime, demonstrating that unmodeled microlensing can produce spurious support for deviations from the Kerr QNM spectrum.
    }
    \label{fig:tgr_bayes_factor}
\end{figure}

We quantify the overall preference for a deviation from the Kerr spectrum using the Bayes factor $\log_{10}\mathcal{B}^{\mathrm{modGR}}_{\mathrm{GR}}$, defined in Eq.~\eqref{eq:logB_modGR_vs_GR}. 
Figure~\ref{fig:tgr_bayes_factor} shows this quantity across the lensing parameter space. 
For the unlensed reference injection, we obtain $\log_{10}\mathcal{B}^{\mathrm{modGR}}_{\mathrm{GR}}=-0.2$, consistent with no preference for the modified-GR hypothesis. 
For the lensed injections, however, the Bayes factor is strongly enhanced for a subset of signals in the wave-optics regime, reaching values as large as $\log_{10}\mathcal{B}^{\mathrm{modGR}}_{\mathrm{GR}}\sim8$.
Although smaller than the overtone Bayes factors, these values still correspond to strong apparent evidence for deviation from GR induced purely by microlensing. 

\subsection{Parameterized analysis using \texttt{pSEOB}}\label{subsec:pseob_results}

\begin{figure}[htbp]
    \centering
    \includegraphics[width=\linewidth]{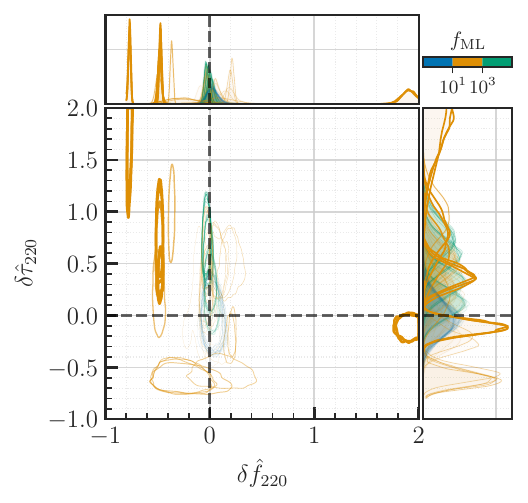}
    \caption{
    Joint $90\%$ credible regions for the fractional deviations in the frequency, $\delta\hat{f}_{220}$, and damping time, $\delta\hat{\tau}_{220}$, of the fundamental $(2,2,0)$ mode obtained from the \texttt{pSEOB} analysis for the $96$ lensed injections. 
    The GR prediction corresponds to $(\delta\hat{f}_{220},\delta\hat{\tau}_{220})=(0,0)$. 
    Significant frequency deviations are concentrated predominantly in the wave-optics regime.
    }
    \label{fig:df_dt_220}
\end{figure}

We perform parameterized analyses using \texttt{pSEOBNRv4HM} waveform model, allowing fractional deviations in the frequency and damping time of the dominant $(2,2,0)$ mode, as seen in Eq.~\eqref{eq:pseob_frac_deviations}.
Figure~\ref{fig:df_dt_220} shows the corresponding $90\%$ joint credible regions for the $96$ lensed injections.
The recovered frequency deviations, $\delta\hat{f}_{220}$, are broadly consistent with the GR prediction in the long-wavelength and geometric-optics regimes, while pronounced departures occur for a subset of injections in the wave-optics regime.
This behavior is consistent with the \texttt{pyRing} results, despite the two analyses using different waveform descriptions and incorporating different portions of the signal. 

The damping-time deviations, $\delta\hat{\tau}_{220}$, exhibit a less coherent pattern across the lensing parameter space. 
As in the \texttt{pyRing} analysis, we therefore do not interpret the damping-time posteriors as a robust diagnostic of the microlensing bias.
Instead, we focus on the frequency deviation, which provides a clearer and more consistent signature across the two complementary analyses.
The filled circles in Fig.~\ref{fig:df_221_220} show the statistical significance of the inferred $\delta\hat{f}_{220}$ deviations. 
The strongest deviations are concentrated in the wave-optics regime, mirroring the behavior obtained for the $(2,2,1)$ frequency in the post-merger analysis. 
Thus, even when the QNM spectrum is constrained simultaneously with the entire signal, unmodeled microlensing can produce an apparent departure from the Kerr prediction for the dominant QNM frequency.

\section{Conclusion}
\label{sec:conclude}

In this work, we investigated the impact of gravitational microlensing on BH ringdown inference and on ringdown-based tests of GR. 
Using GW150914-like lensed injections generated with an isolated point-mass lens, we considered the long-wavelength, wave-optics, and geometric-optics regimes and analyzed the signals using complementary post-merger-only and full IMR waveform models.
Our results demonstrate that unmodeled propagation effects can produce substantial biases in ringdown inference and can mimic additional ringdown signatures that are commonly tested in studies of the Kerr hypothesis.

The strongest effects occur in the wave-optics regime, where frequency-dependent interference produces pronounced amplitude and phase modulations throughout the simulated signal. 
In the ringdown regime, the recovered remnant mass and spin can differ substantially from the corresponding values inferred from the full IMR signal, potentially breaking the consistency between ringdown and IMR measurements. 
By contrast, although the fitting factor can be significantly reduced in the geometric-optics regime, we find comparatively small biases in the recovered ringdown parameters for the injections considered here.
In this limit, the lensed waveform is well described as a superposition of distinct, time-delayed micro-images, each carrying the same underlying QNM spectrum. 
Consequently, an unlensed template may fail to reproduce the post-merger waveform while still recovering QNM frequencies and remnant parameters with relatively small biases. 
This distinction between waveform mismatch and parameter bias is consistent with the behavior found in previous studies~\citep{Mishra:2023vzo,Kopty:2026fnp}.

We further find that microlensing can alter the inferred modal content of the ringdown. 
In particular, wave-optics distortions can produce a strong and spurious preference for the first overtone, with $\log_{10}\mathcal{B}^{221}_{220}\sim40$ in our simulated signals, despite the corresponding unlensed signal showing no preference for the overtone.
In contrast, the evidence for higher-order modes remains essentially unchanged across the lens-parameter space considered here, with $\log_{10}\mathcal{B}^{\rm HM}_{220}$ remaining close to zero.

The same waveform distortions can also produce apparent deviations from the Kerr QNM spectrum. 
For the post-merger-only \texttt{pyRing} analysis, the $(2,2,1)$ frequency can acquire substantial biases in the wave-optics regime, leading to strong apparent preference for the modified-GR hypothesis, with $\log_{10}\mathcal{B}^{\rm modGR}_{\rm GR}\sim8$.
The corresponding damping-time deviations are generally less informative because they are poorly constrained for many injections. 
Importantly, \texttt{pSEOB} analysis shows that microlensing can likewise bias the frequency of the dominant $(2,2,0)$ mode when the QNM spectrum is inferred jointly with the full IMR waveform. 
The consistency between the two analyses indicates that these apparent deviations are not specific to a particular ringdown model and are instead associated with the unmodeled propagation-induced waveform distortions.

The present study is based on a GW150914-like source population, a point-mass lens, zero detector noise, and a restricted range of ringdown models and QNM deviations. 
Extending the analysis to a broader range of binary masses, spins, orientations, and mass ratios, more general lens models, realistic detector noise, and joint inference of the source and lens parameters will be important for quantifying the impact of these effects on individual events and populations. 
Such extensions will be particularly relevant for the higher-SNR ringdown observations expected with current and next-generation detectors, for which increasingly precise measurements of the Kerr spectrum will also increase sensitivity to previously subdominant waveform systematics.
Accounting for propagation-induced distortions in future ringdown analyses will therefore be important for distinguishing genuine departures from the Kerr prediction from apparent deviations generated by gravitational lensing.

\begin{acknowledgments}
    The authors are thankful to Rico~K.~L.~Lo for reviewing the manuscript during the LSC Publications and Presentations procedure and providing very detailed and useful comments.
    AM thanks Prayush Kumar for the helpful discussions and for addressing queries during the initial stages of the project. 
    The authors are grateful for computational resources provided by the LIGO Laboratory and supported by National Science Foundation Grants PHY-0757058 and PHY-0823459. 
    We also acknowledge the use of the IUCAA LDG cluster, Sarathi, where most of the analyses were performed. 
    This material is based upon work supported by NSF's LIGO Laboratory which is a major facility fully funded by the National Science Foundation.
    AM acknowledges the support of the Department of Atomic Energy, Government of India, under project nos. RTI4019 and RTI4013.
\end{acknowledgments}

\software{
    \texttt{Cython}~\citep{cython},
    \texttt{NumPy}~\citep{Harris:2020xlr},
    \texttt{SciPy}~\citep{Virtanen:2019joe},
    \texttt{astropy}~\citep{astropy},
    \texttt{PyCBC}~\citep{pycbc},
    \texttt{LALSuite}~\citep{lalsuite, swiglal},
    \texttt{dynesty}~\citep{Speagle:2019ivv},
    \texttt{BILBY}~\citep{Ashton:2018jfp, Romero-Shaw:2020owr},
    \texttt{corner}~\citep{corner},
    \texttt{Matplotlib}~\citep{Hunter:2007ouj}, and
    \texttt{Jupyter notebook}~\citep{jupyter}.
}

\bibliography{references}{}
\bibliographystyle{aasjournal}

\end{document}